\documentclass[%
 reprint,
 amsmath,amssymb,
 aps,
]{revtex4-1}

\usepackage{graphicx}% Include figure files
\usepackage{dcolumn}% Align table columns on decimal point
\usepackage{bm}% bold math
\usepackage{physics}
\usepackage{xcolor}
\usepackage{wrapfig}

\renewcommand{\[}{\begin{equation}\begin{aligned}} % customise the layout...
\renewcommand{\]}{\end{aligned}\end{equation}}
\newcommand*{\kbt}{k_{\textup{B}}T}
\newcommand*{\kon}{k_{\rm on}}
\begin{document}

\title{Adaptive mechanical proofreading toward faithful clonal selection}


\author{Qing Xu and Shenshen Wang}
\email{shenshen@physics.ucla.edu}
\affiliation{Department of Physics and Astronomy, University of California, Los Angeles, Los Angeles, CA 90095}%Lines break automatically or can be forced with \\

%\affiliation{%

%\affiliation{
%\affiliation{
%\affiliation{%

\date{\today}% It is always \today, today,

\begin{abstract}
To ensure faithful information transmission, cells utilize nonequilibrium drives to reduce errors. Kinetic proofreading is a classic mechanism known to sharpen ligand discrimination by T lymphocytes. It remains an open question whether the adaptive immune system relies solely on kinetic proofreading to boost fidelity. Here, we suggest an alternative: an enhanced form of mechanical proofreading (MPR) in which adaptive force exertion via dynamic cell-cell contact allows faithful selection of high-affinity B lymphocytes. Using a minimal model validated with experiment, we show that adaptive MPR, characterized by mechanical feedback between force generation and contact formation, enables robust discrimination of receptor quality regardless of ligand quantity. Although MPR generically balances the tradeoffs between speed and fidelity, a negative scaling of contact duration with ligand abundance indicates the presence of feedback. Due to its ability to modulate interactions of distinct ligands that share load at membrane contacts, adaptive MPR can be harnessed to mitigate autoimmunity or enhance multivalent vaccines. Overall, this work suggests a generalization of the proofreading mechanism to encompass cellular designs that act across scales to enable competing functionalities.
\end{abstract}

\pacs{Valid PACS appear here}% PACS, the Physics and Astronomy
\maketitle

%\tableofcontents

\section{Introduction}
One basic characteristic of life is the ability to operate with extraordinary fidelity. Key processes responsible for propagating inheritable information inside living cells – protein translation, RNA transcription, and DNA replication – are known to have far lower error rates than expected at thermal equilibrium. It is thus believed that cells utilize non-equilibrium drives to detect and correct errors through a kinetic proofreading mechanism~\cite{ninio1975kinetic, hopfield1974kinetic, ehrenberg1980thermodynamic, murugan2014discriminatory}. This dissipative mechanism reduces errors by amplifying the effect of binding energy differences between competing molecular substrates, via a cascade of enzymatic reactions (proofreading steps) that slow the progress toward the product. Thus, speed is traded for accuracy.

Vital to organismal survival, natural immunity demands efficiency no less than fidelity – in order to contain self-amplifying threats (invasive agents or cancerous cells). Although the ability of T lymphocytes to exquisitely discriminate self from nonself peptides is considered a classic example of kinetic proofreading (KPR)~\cite{mckeithan1995kinetic}, it is unclear if the adaptive immune system relies solely on KPR for fidelity-enhancing purposes.

Here, we suggest an alternative possibility: humoral immunity -- constantly renewing through clonal diversification and selection -- can employ an enhanced form of mechanical proofreading (MPR) to achieve efficient and faithful selection of high-affinity B lymphocytes, producing robust yet tunable responses that are not directly selected for. In contrast to KPR in which the proofreading steps are downstream of receptor binding inside the cell, MPR takes effect through the dynamic junction formed between mutually engaged cells – intracellular forces organize interfacial patterns of force-sensitive receptor-ligand species and modulate information transfer by deforming opposing membranes and molecular complexes. 

Using a minimal construction of force schemes that distills the distinctive feature of MPR,
we contrast schemes that differ in their ability to adapt to the system state. 
We find that, when force can sense and adapt to the contact pattern organized under force, such mechanical feedback 
%between force generation and contact formation 
leads to reduced contact duration with increasing ligand abundance, opposite to when force turns on after a predetermined delay. This feedback allows for robust discrimination of receptor quality against variations in ligand concentration. 
The same feedback causes ligand antagonism, a phenomenon by which non-agonists alter immune responses to co-presented agonists.  
Hence, despite being similar to allostery-based proofreading~\cite{phillips2020molecular} by causing conformational changes in molecular substrates, adaptive MPR is a many-body mechanism in which the manner of force exertion may alter the relationship between molecular species that share load at membrane interfaces. This mechanism has intriguing implications for mitigating autoimmunity and enhancing the efficacy of multivalent vaccines. 

We propose that adaptive MPR at cell-cell contact can enhance the fidelity of clonal selection while meeting competing functional needs, suggesting a generalized proofreading paradigm that encompasses action across scales.

\begin{figure*}[t]
\includegraphics[width=0.98\textwidth]{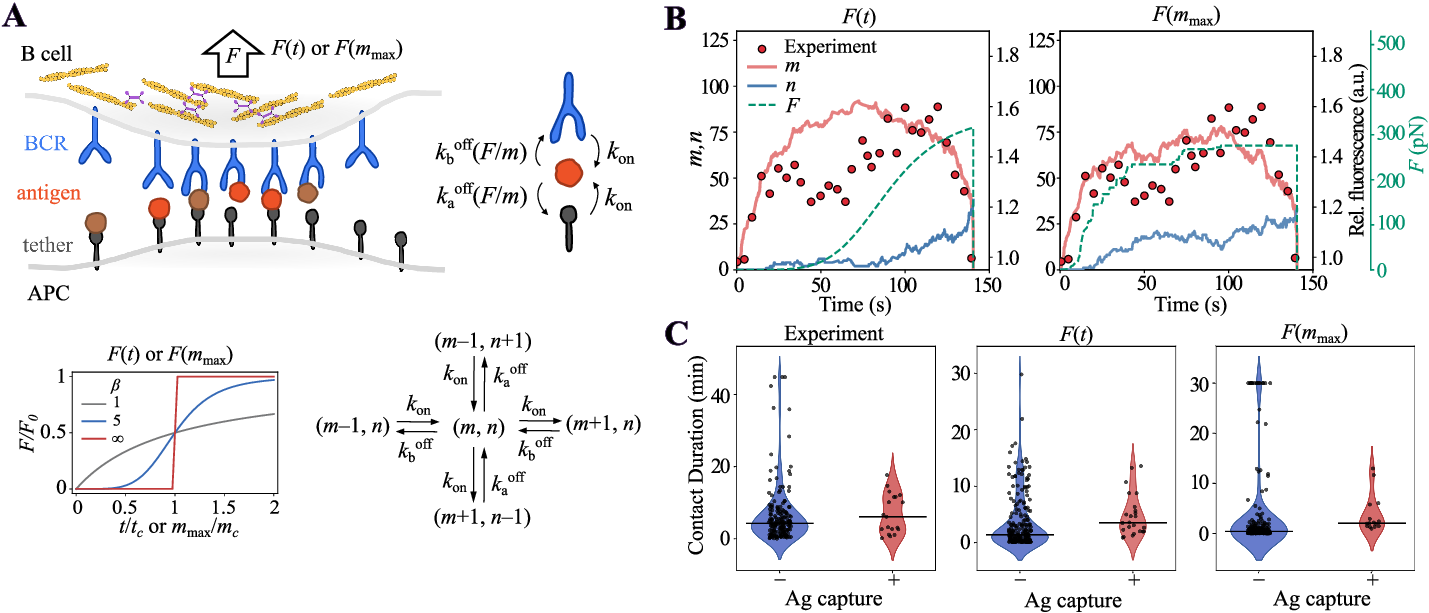}
\caption{\textbf{B cells apply dynamic forces to control antigen extraction: model validation with independent experiments.} 
(A) A B cell uses its BCRs to bind antigens tethered to an APC, forming localized clusters of BCR-Ag-tether complexes. The contractile cytoskeleton exerts dynamic pulling forces on individual clusters and drives antigen extraction. Force influences the system state by altering the off-rates $k_a^\mathrm{off}$ and $k_b^\mathrm{off}$ of the complexes that share load. An inert force $F(t)$ turns on with a preset delay $t_c$, whereas an adaptive force $F(m_{\text{max}})$ compares the maximum cluster size reached, $m_{\text{max}}$, to the onset threshold $m_c$. The system state, specified by the instantaneous cluster size $m$ and the number of BCR-bound antigens $n$, evolves via stochastic binding/unbinding on ether side of the molecular tug-of-wars (Eq.~1).    
%(B) We model these forces as time-dependent (top) or cluster-size dependent (bottom), following a Hill function with parameters $F_0$, $t_c$/$m_c$, and $\beta$ (see Eq.~3).
(B,C) Model captures the observed extraction dynamics. (B) Simulated trajectories of $(m,n)$ and applied force $F$ for each scheme. Cluster size in both schemes (red lines) matches the experimental trend of antigen fluorescence~\cite{natkanski2013} (red symbols).  (C) Distributions of contact duration  for low/high extraction B cells (“−”/“+”) in both schemes (right two panels) match experimental data~\cite{suzuki2009} (left panel). High-extraction B cells show a larger median (higher black bar), whereas lower-extraction B cells are skewed toward brief contacts with a tail toward long durations. 
Parameters: $L_0 = 100$, $k_{\text{on}} = 0.05s^{-1}$, $E_a = 12.6k_BT$, $E_b = 13.3k_BT$, $x_a = 1.5 \text{nm}$, $x_b = 2 \text{nm}$, $\beta = 5$; (B) $F_0 = 350 \text{pN}$, $t_c = 1.5 \text{min}$, $m_c = 60$; (C) 
%n_{\text{Ag}}^{\text{threshold}} = 42$, $\text{\# B cells}=250$, 
middle: $F_0/\text{pN} \sim \mathcal{U}(250, 900)$, $t_c/\text{min} \sim \text{LogUniform}(0.1, 15)$; right: $F_0/\text{pN} \sim \mathcal{U}(300, 900)$, $m_c \sim \mathcal{U}(20,80)$.
}
\label{cluster}
\end{figure*}

\section{Model}

Within the transient microenvironment called germinal centers (GCs), B lymphocytes improve their antigen recognition capability through a rapid Darwinian process known as affinity maturation~\cite{eisen1964variations, victora2012, victora2022germinal}. In a cyclic fashion, GC B cells diversify their receptor encoding genes during replication via somatic hypermutation and compete for limited survival signals provided by T helper cells. Ideally, B cells expressing higher affinity receptors (BCRs) compete better and expand preferentially. 

The key step that translates the affinity excess of a B cell clone to its competitive advantage is a vigorous physical process of antigen extraction~\cite{batista2001, suzuki2009, natkanski2013, mcarthur2025antigen}: through a dynamically organized intercellular junction (immunological synapse), a B cell actively acquires antigens from an antigen presenting cell (APC) using cytoskeletal forces and presents the acquired antigens in a processed form on its surface. T helper cells then discern BCR affinity according to the surface density of presented antigens~\cite{lanzavecchia1985antigen, victora2010germinal}. Consequently, antigen extraction dynamics – tunable through force exertion – control, and potentially limit, the fidelity of clonal selection.

Upon binding of a BCR to an antigen (Ag) tethered to the surface of an APC, a BCR-Ag-tether 3-body complex forms. Productive binding deforms the B-cell membrane and drives clustering of BCRs, initiating a series of BCR-Ag interactions. These interactions in turn trigger cytoskeletal remodeling inside the B cell and generation of contractile forces needed to pull Ags away from the APC. Successful Ag extraction occurs when the BCR-Ag bond withstands the forces longer than the Ag-tether bond within a 3-body complex. Once all complexes in a cluster are disrupted, the B cell locally detaches from the APC, concluding one extraction attempt. 

Previous models~\cite{jiang2023immune, jiang2023molecular, jiang2024physical} assume pre-formed clusters subject to constant loads, only depicting the rupture stage. Here, we describe the entire process of an extraction attempt – from initial binding and subsequent cluster growth to complete dissociation – under the influence of dynamic forces throughout, as demonstrated in experiments combining live-cell imaging and sensitive force measurements~\cite{Nowosad:2016, Wang:2018}. This allows us to explore the functional role of mechanical feedback through which force not only alters the system state but can adjust to the altered state. Such feedback was shown~\cite{knevzevic2018active} to be able to stabilize a multifocal synaptic pattern comprising segregated BCR-Ag clusters, characteristic of GC B cells undergoing affinity maturation~\cite{Nowosad:2016}. We thus focus on antigen extraction via one localized cluster (Fig.~1A).

\textit{Antigen extraction dynamics.} Assume that a limited amount $L_0$ of antigens is available along with a large abundance of BCRs; both are uniformly distributed without spatial specificity. Changes in the quantities of the BCR-Ag-tether complexes ($m$), BCR-Ag complexes ($n$) and Ag-tether complexes ($L_0-m-n$) resemble the turnover of chemical species due to stochastic reactions (here, binding/unbinding events). At each moment, the pair $(m, n)$ fully specifies the state of the system, where $m$ represents the cluster size and $n$ the number of BCR-bound Ags; the value of $n$ at final dissociation corresponds to the amount of extracted antigen, $n_\mathrm{Ag}$. The probability of occupying the state $(m,n)$ evolves according to a set of one-step master equations:
\[
\dv{P(m, n; t)}{t}=&\Big[(\xi^{1, -1}-1)mk_{a} +(\xi^{1, 0}-1)mk_{b}  + \\
(\xi^{-1, 1}-1)n \kon
&+(\xi^{-1, 0}-1)(L_0-m-n)\kon \Big]P(m, n).
\label{master_eq}
\]
Here $\xi^{i, j}$ is the step operator defined through $\xi^{i, j}G(m, n) = G(m+i, n+j)$ for any function $G(m, n)$. Note that $0\leq m+n \leq L_0$. The first two terms on the right hand side of Eq.~\ref{master_eq} represent breaking of the 3-body complex at the Ag-tether interface and the BCR-Ag interface, respectively. The latter two terms correspond to the reverse association processes. Sample trajectories of $(m, n)$ are shown in Fig.~1B (red and blue lines). 

Ag acquisition assumes a tug-of-war architecture~\cite{jiang2023molecular}, in which kinetics of competitive bond rupture between the tethering (Ag-tether) and tugging (BCR-Ag) interactions govern the likelihood and speed of antigen extraction~\cite{jiang2023immune}. 
We assume that two binding interfaces share the same constant on-rate $k_{\rm on}$ but have different off-rates (Fig.~1A). These off-rates depend on affinity and vary with the applied force according to Bell's phenomenological model that describes slip bonds~\cite{bell1978}: 
\[
&k_a^{\text{off}}(m)=k_{0}e^{ -(E_a -\frac{Fx_a}{m})/\kbt}\\  &k_b^{\text{off}}(m)=k_{0}e^{-(E_b-\frac{Fx_b}{m})/\kbt}
\label{rates}
\]
Here $F$ is the total force exerted on the cluster that we set to be dynamic (see below). The evenly distributed force per bond $F/m$ is key to extraction dynamics. $E_a$ and $E_b$ are binding free energies of the Ag-tether and BCR-Ag interactions, while $x_a$ and $x_b$ are the corresponding bond lengths, each indicating the distance between the potential minimum and the energy barrier to bond rupture along the reaction coordinate. %\cite{Erdmann:2004b}.
%In this study we fix $x_a=1.5$nm, and $x_b=2.0$nm.  
$k_0$ is the basal rate of dissociation, aggregating the effect of additional factors that are not explicitly modeled, including membrane fluctuations and hydrodynamics.  %$k_{a0}=k_0 
%The full simulation details are given in SI (see Sec.2).

\paragraph*{Dynamic force schemes.} 
Cytoskeletal forces may sense cluster size (though size-dependent assembly of the pulling machinery) or turn on after a predetermined delay (via a reaction cascade of signal propagation). A minimal construction of force schemes that encodes the essence of inert and adaptive MPR is given by  
%Explicitly, we consider time-dependent dynamic force and cluster size-dependent force, which are assumed to take the following forms
\[
F(t) =F_0 \frac{t^\beta}{t^\beta+t_c^\beta}, \,\, \,\,
F(m_{\max}) = F_0\frac{m_{\max}^\beta}{m_{\max}^\beta+m_c^\beta}.
\]
Both schemes are three-parameter families with tunable magnitude ($F_0$), nonlinearity ($\beta$) and threshold of onset ($t_c$ or $m_c$). The major difference is whether the force adapts to system state: inert force $F(t)$ rises to the half-maximum magnitude after a fixed lag $t_c$, whereas adaptive force $F(m_{\max})$ adjusts to the maximum cluster size $m_{\max}(t) = \max_{t'\leq t}[m(t')]$ ever reached until a given time $t$. This choice reflects the ratchet-like nature of the contractility apparatus~\cite{mason2011tuning, komianos2018stochastic}, such that force scales with the maximum cluster size so far accessed and is insensitive to instantaneous fluctuations. %(like observed in ~\cite{tolar2014force, wang2018}). 
By exploring the parameter space of both force schemes, we will identify the dynamic regimes accessible to the extraction system and determine key functional consequences of adaptive force application.

\section{Results}
Immune cells appear to be able to fulfill seemingly incompatible functional properties: speed of antigen detection, fidelity of affinity discrimination, and robustness to variations in antigen quantity. We will show that the exertion of adaptively controlled tugging forces might provide a common solution to disparate objectives. Our model thus gives a window on how B cells achieve selective expansion of high-quality clones while balancing competing needs.

\begin{figure}[t]
\includegraphics[width=0.3\textwidth]{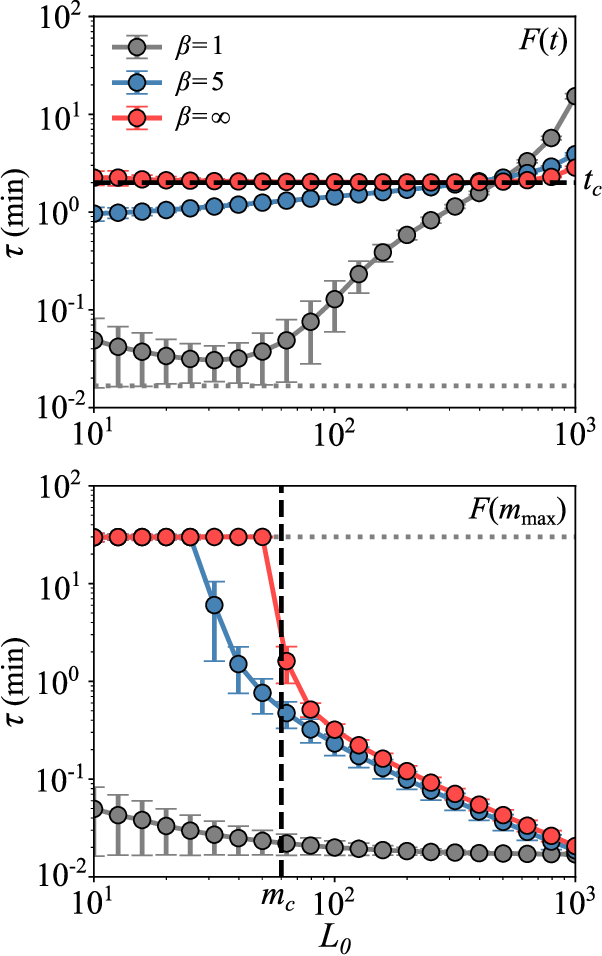}
\caption{\textbf{Inert and adaptive forces produce opposite scaling of contact duration with antigen quantity.}  At each antigen quantity $L_0$, contact duration $\tau$ (i.e. cluster lifetime) is averaged over 1000 independent runs. $\tau$ is limited to a realistic range [$t_\text{min}$, $t_\text{max}$] as indicated by the dotted lines. 
%The top panel is for a time-dependent force $F(t)$, and the bottom panel is for a cluster-sensing force $F(m_{\text{max}})$.  
Under $F(t)$, contact typically prolongs as the amount of presented antigens increases. Under $F(m_{\text{max}})$, contact duration decreases with increasing antigen abundance.  $k_{\text{on}} = 0.05\text{s}^{-1}$, $E_a = 12.6k_BT$, $E_b = 15k_BT$, $F_0 = 4000\text{pN}$, $t_c = 2 \text{min}$, $m_c = 60$, $t_\text{min}=1 \text{s}$, $t_\text{max}=30 \text{min}$.
}
\label{duration}
\end{figure}

\subsection{Model captures observed extraction dynamics}
Significant experimental progress has been made over the past few decades in characterizing the interaction between B cells and APCs and visualizing subsequent antigen extraction and internalization~\cite{batista2000b,batista2001,allen2007,suzuki2009,natkanski2013,nowosad2016,kwak2018}. In a pair of early works~\cite{batista2000b, batista2001}, Batista and Neuberger 
%showed that B cells can extract antigens tethered to a surface and 
proposed that antigen extraction occurs via exertion of mechanical forces. Direct evidence has recently been provided in studies showing that B cells physically pull on BCR-Ag-tether complexes through synaptic contacts, leading to affinity-dependent acquisition of BCR-Ag microclusters~\cite{natkanski2013,nowosad2016,spillane2017b}.

Our model largely recapitulates the observed extraction dynamics (see simulation details in SI). As shown in Fig.~1B, the simulated trajectory of cluster size (red line) closely matches the temporal data of antigen fluorescence (red dot)~\cite{natkanski2013} both in the initial stage of steady cluster growth and at the final stage toward complete dissociation -- when force becomes strong enough to trigger a cascade of bond rupture. This cascade reflects load-sharing among 3-body complexes: as more bonds break, the remaining complexes bear a larger force per bond, hence an accelerated dissociation. Dynamics at intermediate times are more complex, with non-monotonic variations in cluster size potentially due to pulsatile contractility of the cytoskeleton.
Note that despite distinct force trajectories (green dashed lines), both schemes can capture the essential characteristics of cluster formation and dissociation. 

An independent validation can be obtained from the conditional distribution of contact duration. To seek correlations between contact duration and antigen capture among B cells, Suzuki et al.~\cite{suzuki2009} sorted the measured duration based on whether or not a detectable amount of Ag was acquired (Fig.~1C, each symbol being a B cell). Both force schemes reproduce the key features of each group: The negative group is strongly skewed toward short durations below 2 minutes, while having a long tail that extends beyond 30 minutes. In contrast, the positive group shows a narrower range with a higher median (black bar). Importantly, capturing the wide span of contact duration requires sufficient heterogeneity in force parameters among cells (see SI and Fig.~S2): the long tail corresponds to small magnitudes and large onset thresholds, and the opposite accounts for brief contacts. The limited range in the positive group reflects that productive Ag extraction imposes constraints on $F_0$, such that it is neither too high to allow cluster growth nor too low to trigger dissociation. 
%Therefore, extraction trajectories and conditional contact duration provide independent validations of our minimal model. 

The speed of cluster extraction is crucially dependent on the timing at which the rupture cascade is triggered. It is thus useful to ask whether a tipping point underlies this loss of stability, or it is merely a result of stochasticity in binding/unbinding kinetics. A bifurcation analysis of the deterministic dynamics indicates that, as the total force $F$ increases, two steady states, $m_s^\mathrm{low}$ and $m_s^\mathrm{high}$, get closer and eventually merge and annihilate at a tipping point $(F^*, m^*)$, beyond which cluster size vanishes (Fig.~S1). The critical value $F^*$
%$F^*\approx L_0\min\left(f_a \rm{p}\ln\left(k_{\mathrm on}/e k_a^\mathrm{off}\right), f_b \rm{p}\ln\left(k_\mathrm{on}/e k_b^\mathrm{off}\right)\right)$ 
is determined by the weaker side of the tug of war (see SI for expressions).
%here $f_a=k_B T/x_a$ and $f_b=k_B T/x_b$ are rupture forces driven by thermal noise and the product logarithm $\rm{p}\ln(a)$ is evaluated as the solution $x$ to $xe^x=a$. Perturbation analysis shows that 
Since $m_s^\mathrm{high}$ is stable whereas $m_s^\mathrm{low}$ is not, if the cluster and force develop gradually, the system would approach $m_s^\mathrm{high}$ and fluctuate around it until force exceeds $F^*$ triggering dissociation. If, instead, $m$ falls below $m_s^\mathrm{low}$ early on, the rapid increase in force per bond would cause the cluster to dissolve immediately. This bifurcation underlies the constraints on $F_0$ described above.
%These distinct behaviors are indeed what we saw when overlaying stochastic trajectories onto the bifurcation diagram (Fig.~S).

\begin{figure*}[t]
\includegraphics[width=0.96\textwidth]{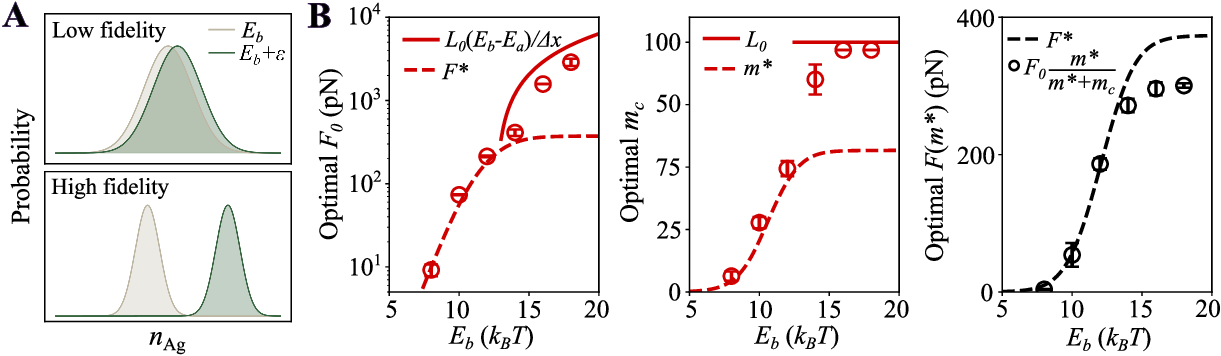}
\caption{\textbf{Optimal force scheme for faithful selection depends on BCR affinity.}  (A) Selection fidelity is determined by distinguishability of readout (here, extraction level $n_\mathrm{Ag}$) distributions between similar BCR affinities ($\epsilon\ll E_b$).
%Improved B cell affinity discrimination occurs when the probability distributions of extracted antigen amounts at nearby BCR affinities are more distinct. Ranking fidelity quantifies affinity discrimination performance.  
(B) Optimal parameters $F_0$ and $m_c$  of an adaptive force that maximizes the ranking fidelity $\xi$ (shown by symbols) vary with BCR affinity. At low $E_b$, they correspond to the bifurcation point (dashed line). At high $E_b$, they close the affinity gap under a step force (solid line).  
The left two panels (red) are for step force $F(m_{\text{max}}; \beta = \infty)$ and the right panel (black) is for linear force $F(m_{\text{max}}; \beta = 1)$. 
Parameters: $k_{\text{on}} = 0.05\text{s}^{-1}$, $L_0 = 100$, $E_a = 12.6k_BT$, $\epsilon = 0.5k_BT$, $x_a = 1.5 \text{nm}$, $x_b = 2 \text{nm}$.
}
\label{fidelity}
\end{figure*}

Then, what measurement can distinguish the underlying force scheme? Interestingly, our model predicts that the scaling relationship between the mean contact duration $\tau$ and ligand quantity $L_0$ can tell apart inert and adaptive forces (Fig.~2). Under $F(t)$, a larger abundance of ligand leads to a longer contact, whereas under $F(m_\mathrm{max})$, the extraction proceeds faster as the amount of ligand increases -- a desirable feature for timely containment of insults. 
%[There is experimental evidence supporting the latter: Omer ppr] 
Moreover, the nonlinearity $\beta$ of force onset influences the sensitivity of $\tau$ to changes in $L_0$ depending on the force scheme; stronger nonlinearity increases the sensitivity under $F(m_\mathrm{max})$ but decreases the sensitivity under $F(t)$. These differences reflect distinct conditions to be met in order to dissociate the cluster (see SI for details). %Analytical intuition can be obtained in certain cases (see SI): 
Under $F(t)$, the contact duration is set by the time taken to reach the bifurcation point. %($F^* < F_0$). 
That is, $F(\tau)=F^*\propto L_0$, leading to
\[
\tau \sim L_0^{1/\beta}.
\]
Under $F(m_{\mathrm max})$, for a step force ($\beta=\infty$), the contact lasts until the cluster grows to the critical size that triggers force, i.e., $m(\tau)=m_c$, resulting in
\[
\tau \sim L_0^{-1}.
\]
These predictions agree well with simulations (Fig.~2) and could be tested by measuring the antigen-dose dependence of the mean duration of contact. Such an experiment will inform the dynamic characteristics of force that B cells employ to probe their receptor quality.

\subsection{Optimal force scheme for faithful clonal selection}
%Within this cluster formation-disassociation process, many quantities are affinity dependent and thereby can potentially be used as affinity readout for discrimination. In this section we investigate possible readouts (maximal cluster size, contact (signaling) duration, and antigen acquisition. We propose that the switching between readouts is necessary to achieve the observed wide-range affinity discrimination. 
Efficient selection of high-affinity B cell clones relies on a faithful mapping from an excess in BCR affinity to a competitive advantage in acquiring limited T-cell help~\cite{gitlin2014clonal}. T-help acquisition is determined by the amount of antigen presented on the B cell surface~\cite{lanzavecchia1985antigen, victora2010germinal} and, in turn, by the extraction level $n_\mathrm{Ag}$. Stochastic extraction dynamics yield noisy affinity readouts (here, $n_\mathrm{Ag}$) and hence unreliable clonal selection. 
We therefore quantify the fidelity of clonal selection, $\xi$, as the probability that a B cell with a higher affinity $E_b+\epsilon$ extracts a larger amount of antigen than a peer with affinity $E_b$, representing a correct ranking:
\[
\xi= {\rm Prob}[n_{\rm Ag}(E_b+\epsilon) > n_{\rm Ag}(E_b)].
\]

%The stochastic nature of the extraction process, in a large part, makes the mapping from affinity to extraction level unreliable. 
As such, selection fidelity is limited by the distinguishability between readout distributions with similar affinities~\cite{jiang2024physical}, as illustrated in Fig.~3A. At one extreme, if two distributions do not overlap, then $\xi\approx 1$; at the other extreme, with two identical distributions, $\xi=0.5$, indicating that cells drawn from either distribution have an equal chance of extracting more antigen and being selected. Often the distribution of $n_\mathrm{Ag}$ is not analytically tractable, then $\xi$ can be estimated from the rate of correct ranking with many simulated B cell pairs (with $E_b$ vs. $E_b+\epsilon$).
In cases where $n_\mathrm{Ag}$ follows a Gaussian distribution, with mean $\mu_n$ and variance $\sigma_n^2$ both dependent on affinity, the fidelity to the leading order in $\epsilon$ (i.e. in the hard discrimination regime) is given by
\[
\xi \approx \frac{1}{2} + \frac{1}{2\sqrt{\pi}\sigma_n}\dv{\mu_n}{E_b}\epsilon.
\label{xi_approx}
\] 
It follows that selection is faithful when the mean extraction level is sensitive to changes in affinity (large $\mathrm{d}\mu_n/\mathrm{d}E_b$) and/or the variance in extraction is low (small $\sigma_n$). This allows us to understand how forcing dynamics influence selection fidelity through their impact on the moments of the extraction level distribution, $P(n_{\rm Ag})$. 

\begin{figure*}[t]
\includegraphics[width=0.97\linewidth]{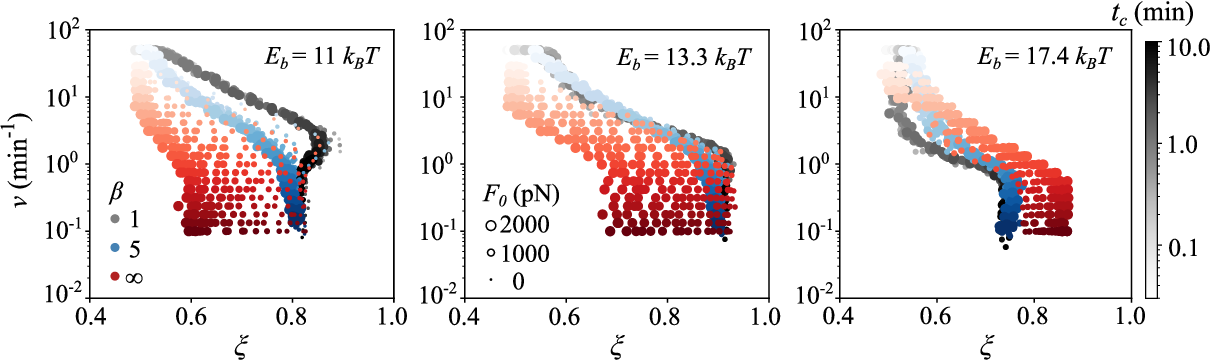}
\caption{\textbf{Speed-fidelity trade-off in BCR-affinity discrimination based on antigen extraction.}  Extraction speed and ranking fidelity are displayed simultaneously for each set of force parameters $\beta$, $F_0$, and $t_c$, represented by symbol color, size, and shading, respectively. Each symbol is an average over 4000 replicate simulations. Three panels correspond to different BCR affinities, $E_b=11k_BT$, $13.3k_BT$, and $17.4k_BT$ from left to right. At low $E_b$, a linear force (in grey) defines the Pareto front, whereas at high $E_b$, a step force (in red) optimally balances the trade-off. A further delay beyond the turning point (the elbow) slows extraction without gain in fidelity.  Parameters: $L_0 = 100$, $k_{\text{on}} = 0.05\text{s}^{-1}$, $E_a = 12.6k_BT$, $\epsilon = 0.5k_BT$. An inert force $F(t)$ is used here. An adaptive force $F(m_{\text{max}})$ yields similar behaviors (see Fig.~S5).
 }
\label{spd_acc_tradeoff}
\end{figure*}

To quantify the contribution from different stages of the extraction process, we express the distribution as follows:
\[
P(n_{\rm Ag}) = \sum_{m_{\rm tot}=0}^{L_0}P(m_{\rm tot})P(n_{\rm Ag}|m_{\rm tot}). 
\label{Prob_nag}
\]
Here the distribution of $m_\mathrm{tot}$, the total number of distinct antigens visited by BCRs in one extraction attempt, encodes the dynamic accessibility of antigens on the APC and characterizes the stage of cluster formation. The conditional probability $P(n_{\rm Ag}|m_{\rm tot})$ is governed by the rupture stage in which antigen extraction occurs with the following success chance (treating rupture of the BCR bond and the tether bond as independent):
\[
\eta(f_r) = \frac{k_{a}^{\rm off}}{k_{a}^{\rm off}+k_{b}^{\rm off}} = \frac{1}{1+e^{-(E_b-E_a-f_r(x_b-x_a))/\kbt}}.
\]
Note that $\eta$ depends on the force per bond $f_r$ just before rupture and $f_r$, in turn, depends on total force and the number of remaining 3-body complexes, both of which change over time. Therefore, antigen extraction can be viewed as a series of $m_\mathrm{tot}$ Bernouli events with varying success chances $\eta$ due to varying rupture forces (see SI), leading to simplified expressions of the moments: 
\[
&\mu_n =\langle m_{\rm tot} \overline{\eta} \rangle\approx \langle m_{\rm tot}\rangle \langle\overline{\eta}\rangle, %= m_{\rm tot}\int \eta(f)P(f)df\\
\\
&\sigma_n^2 \approx \langle m_{\rm tot}\rangle  \langle \overline{\eta} -\overline{\eta^2} \rangle.
\label{mean_std}
\]
Here, $\langle \cdot \rangle$ represents an ensemble average (over stochastic binding/unbinding events) and $\overline{\eta^\alpha} = \int_0^{\infty}\eta^{\alpha}(f_r)P(f_r)df_r$ denotes the $\alpha$th moment of the extraction chance, averaged over the distribution $P(f_r)$ of rupture force per bond. A relatively narrow distribution of $m_{\rm tot}$ factorizes the ensemble average. 
%$m_\mathrm{tot}$ and $\langle \bar{\eta} \rangle$ can in general be obtained from simulated trajectories, while in special cases, such as with step forces, Eq.~\ref{mean_std} can be evaluated analytically. Considering the dominant factor $\rm{d}\mu/\rm{d}E_b$ in the expression of $\xi$ implies that
Clearly, both cluster formation and dissociation contribute to the sensitivity
\[
\dv{\mu_n}{E_b} \approx \langle\overline{\eta}\rangle \dv{\langle m_{\rm tot}\rangle }{E_b} + \langle m_{\rm tot}\rangle \dv{\langle \overline\eta\rangle }{E_b},
\label{SNR}
\]
reflecting that B cells of different affinities engage with different amounts of antigen during cluster growth (first term) and dissociate at disparate extraction chances (second term). Two stages dominate the contribution at low and high affinities, respectively (see Fig.~S3). 

We can now identify the dynamic force scheme that enhances affinity discrimination to optimize selection fidelity. By varying the force magnitude $F_0$ and the onset threshold $m_c$ (or $t_c$) under different nonlinearities $\beta$, we determine the highest fidelity achievable at various BCR affinities. A notable switch in behavior occurs (Fig.~3B): at high BCR affinities ($E_b>E_a$), a step force ($\beta=\infty$) of magnitude $F_0 \sim L_0(E_b-E_a)/\Delta x$ yields the optimal fidelity (red solid lines), whereas at low BCR affinities ($E_b<E_a$), fidelity is optimal when $F(m^*)=F^*$ is satisfied (dashed lines). Below we provide some physical intuition. 

At high BCR affinities, 
%a step force is required to generate sufficient rupture force to maximize sensitivity in antigen extraction, $\mathrm{d}\eta/\mathrm{d}E_b$. When $E_b\gg E_a$, 
the number of antigens visited is primarily limited by $E_a$ (the weaker bond in the three-body complex), resulting in low sensitivity of $m_\mathrm{tot}$ to $E_b$ during cluster formation. Hence, fidelity at high $E_b$ is mainly determined by the rupture stage (see Fig.~S3C). To optimize fidelity, both a large $m_{\mathrm{tot}}$ and a high $\mathrm{d}\eta/\mathrm{d}E_b$ are desired. A large $m_\mathrm{tot}$ can be obtained via a step force with a large onset threshold $m_c$ (or a large $t_c$), allowing time for cluster growth. $\mathrm{d}\eta/\mathrm{d}E_b$ reaches its maximum when $k_b^{\mathrm{off}} = k_a^{\mathrm{off}}$~\cite{jiang2024physical}, which occurs when the rupture force closes the affinity gap between the BCR bond and the tether bond:
\begin{equation}
F\Delta x/m\sim E_b-E_a,
\label{F_optimal}
\end{equation} 
where $m$ is the instantaneous cluster size. The optimal $F_0$ and $m_c$ found in simulations indeed match this gap-closing condition (red solid lines in Fig.~3B and Fig.~S4A).

At low BCR affinities, a force that regulates cluster growth is key to optimizing fidelity. The optimal scheme is found to meet a general criterion: 
\[ 
F(m^*) = F_0 \frac{(m^*)^{\beta}}{(m^*)^{\beta}+m_c^{\beta}} = F^*, 
\label{F_m}
\]
as shown by dashed lines in Fig.~3B (and blue line in Fig.~S4A). This condition results from the strong sensitivity of cluster formation, i.e. large $\mathrm{d} m_\mathrm{tot}/\mathrm{d}E_b$, near the bifurcation point $(F^*, m^*)$, where a small increment in $E_b$ significantly alters system dynamics (see Fig.~S3B). An optimal inert force $F(t)$ shows a similar trend of dependence of $F_0$ and $t_c$ on $E_b$ (comparing Fig.~S4B to Fig.~3B). 

Although it is not immediately clear how B cells infer the exact force that closes the affinity gap or matches the tipping point, our results suggest a compelling possibility: B cells may switch the force application strategy (hence discriminatory regime) as they evolve from below to above $E_a$ during affinity maturation --- potentially through sensing the tether strength.                 
\begin{figure*}[t]
\begin{center}
\includegraphics[width=\textwidth]{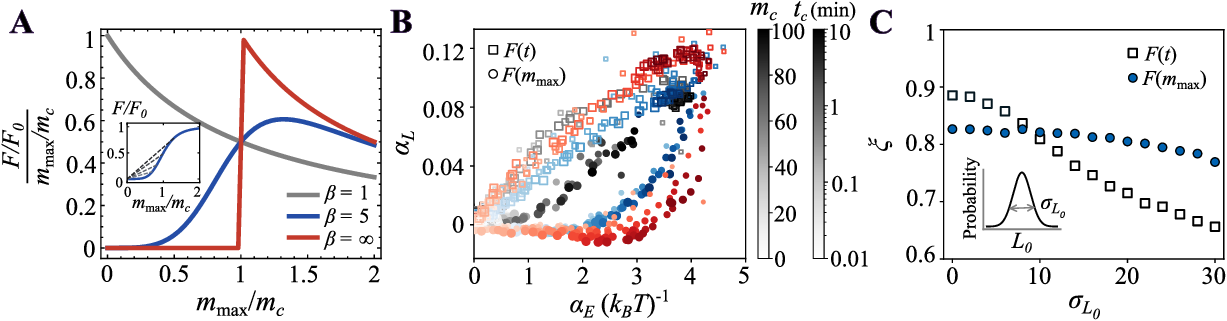}
\caption{\textbf{Absolute discrimination can be achieved with non-linear adaptive force.}  (A) Scaled force per bond is plotted against scaled maximum cluster size according to the definition of an adaptive force (Eq.~3). For $\beta > 1$, force per bond increases with maximum cluster size before the total force saturates, providing a negative feedback that limits cluster growth. The inset shows $F(m_{\text{max}}, \beta = 5)$, where the slope of the dash lines represents force per bond.  (B) Sensitivity of mean extraction level to changes in ligand quantity, $\alpha_L$, against sensitivity to changes in BCR affinity, $\alpha_E$ (Eq.~14). Force parameters $\beta$, $F_0$, and $t_c$ (or $m_c$) are represented by symbol color, size, and shading, respectively. $\alpha_L$ and $\alpha_E$ are averaged over 5000 independent runs for each set of force parameters. Under $F(t)$, $\alpha_L$ always increases with $\alpha_E$. Under $F(m_{\text{max}})$, especially with a strong non-linearity, there is a regime where $\alpha_E$ reaches larger values while $\alpha_L$ stays near zero, i.e., absolute discrimination.  
(C) Ranking fidelity is calculated for 6000 pairs of B cells, each seeing a different ligand number sampled from a Gaussian distribution with mean $L_0$ and standard deviation $\sigma_{L_0}$. $F(m_{\text{max}})$ allows robust discrimination against strong variations in ligand number.  Parameters: (B, C) $L_0=100$, $k_{\text{on}} = 0.05\text{s}^{-1}$, $E_a = 12.6k_BT$, $E_b = 13.3k_BT$, (B) $F_0 = 100$--$1000$pN with a spacing of $100\text{pN}$, (C) $F_0 = 800\text{pN}$, $t_c = 1\text{min}$, $m_c = 60$, $\beta = 5$.
}
\label{absdis}
\end{center}
\end{figure*}

%\subsection{Discrimination based on antigen acquisition displays speed and fidelity trade-off}
\subsection{Speed-fidelity tradeoff in extraction-based affinity discrimination}

Intuitive speed-accuracy tradeoffs have been reported for various information-processing systems that involve kinetic proofreading (e.g., ~\cite{wang2021theory,cui2018identifying}). Does mechanical proofreading based on active, physical signal acquisition presented here experience similar tradeoffs? 

Tradeoffs manifest as a Pareto front along which one trait cannot improve without deteriorating the other~\cite{shoval2012evolutionary}. To reveal the Pareto front, we explore the parameter space of dynamic forces and present the speed and accuracy of many ``force variants” over a wide dynamic range. We use the inverse  cluster lifetime (from initial contact to dissociation) to represent speed and identify accuracy with selection fidelity. As shown in Fig.~4, for each given nonlinearity $\beta$, the Pareto front is associated with a particular force magnitude $F_0$; varying the onset threshold $t_c$ (or $m_c$) moves the system along the front.  

Two distinctive features draw our attention. First, the shape and nature of the Pareto front depend on affinity. While a linear force ($\beta=1$, gray symbol) yields the best performance at low affinities, a step force ($\beta=\infty$, red symbol) performs favorably at high affinities. For the former, a linear force restricts low-affinity BCRs from clustering but allows higher-affinity B cells to form substantial clusters prior to extraction, facilitating rapid discrimination. The latter occurs because at high $E_b$, a step force supports fast cluster growth and can confer higher fidelity for a given $m_\mathrm{tot}$ (by closing the affinity gap). At an intermediate $E_b$, the fronts at different $\beta$ coincide.
Second, an optimal balance is achieved at an intermediate delay of force onset --- further delays yield no more fidelity gain but significantly slow extraction. At low $E_b$, there is a parameter region where speed and fidelity improve simultaneously.
%(especially clear under a linear force). 
The turning point (the elbow) corresponds to the fidelity reaching its maximum due to strong sensitivity at the tipping point. At high $E_b$, the turning point marks the condition under which a later force onset reduces speed without improving fidelity.

In sum, the turning point in the speed-fidelity Pareto front designates the optimal force dynamics that best enable rapid and faithful discrimination of BCR affinity through antigen extraction. 

\subsection{Absolute discrimination under adaptive force}
%\subsection{Absolute discrimination under cluster sensing force}
Another desirable feature of affinity discrimination is robustness against variations in ligand concentration. For T lymphocytes, a proofreading cascade supplemented by a negative feedback (mediated by the SHP-1 phosphatase) was found to enable detection of a few foreign ligands against a large excess of self-ligands \cite{altan2005modeling, feinerman2008, Francois:2013}. 
This ability to distinguish ligand quality regardless of their quantity was named absolute discrimination \cite{Francois:2016a}. Analogously, as GC B cells migrate through the follicular dendritic cell network to sample heterogeneously distributed antigens \cite{suzuki2009, martinez2023long}, the amount of extracted antigen should reflect BCR quality irrespective of encountered antigen concentration. Can B cells achieve absolute discrimination using mechanical control?

\begin{figure*}[t]
\begin{center}
\includegraphics[width=0.95\textwidth]{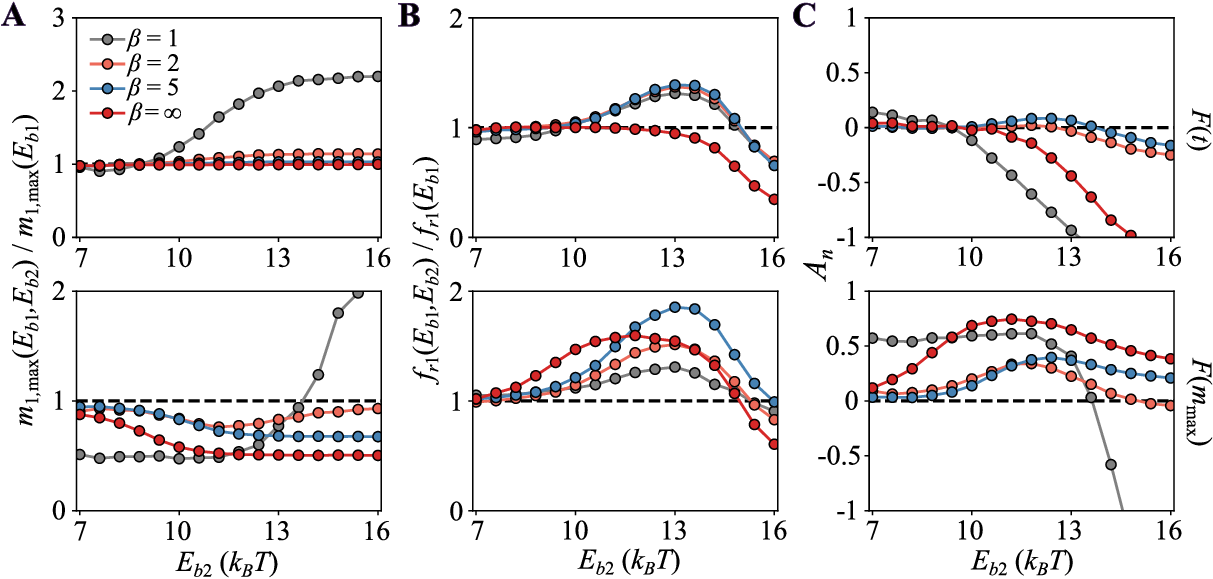}
\caption{\textbf{Antagonism due to load sharing in the presence of a second antigen type.}  The ratio of the maximum agonist cluster size $m_{1,\text{max}}$ (A) and that of the mean rupture force per agonist bond $f_{r1}$ (B), between cases with and without antagonists, combine to determine the strength of antagonism $A_n$ (C), as defined in Eq.~15. The top/bottom row corresponds to an inert/adaptive force. All results are averaged over 1000 replicate simulations.  
Under $F(t)$, mixed antigens are largely cooperative ($A_n < 0$) due to larger agonist clusters attained. Under $F(m_{\text{max}})$, mixed antigens are mostly antagonistic ($A_n > 0$), caused by restricted growth of agonist clusters combined with stronger rupture force per agonist bond that lowers the chance of successful antigen extraction. $L_{10} = L_{20} = 40$, $k_{\text{on}} = 0.05\text{s}^{-1}$, $E_a = 12.6k_BT$, $E_{b1} = 15k_BT$. Top: $F_0 = 700\text{pN}$, $t_c = 1\text{min}$. Bottom: $F_0 = 500\text{pN}$, $m_c = 40$.
% // simple, focused point
% // no unnecessary information
% // formation 
% // intermediate regime
}
\label{antagonism}
\end{center}
\end{figure*}

To probe conditions under which this can be achieved via Ag extraction, we vary the force parameters and compute the sensitivity of the extraction level to small changes in receptor affinity ($\alpha_E$) and antigen quantity ($\alpha_L$), respectively:
\[
\alpha_E = \frac{1}{\sigma_n}\dv{\mu_{n}}{E_b}, \qquad \alpha_L = \frac{1}{\sigma_n}\dv{\mu_n}{L_0}.
\]
Interestingly, as shown in Fig.~5B, only nonlinear adaptive forces ($F(m_\mathrm{max})$, red and blue circles) produce a regime of absolute discrimination, where $\alpha_L$ stays near zero as $\alpha_E$ reaches a large value. In contrast, inert forces indifferent to clustering state ($F(t)$, squares) cannot decouple receptor quality from ligand quantity; $\alpha_E$ and $\alpha_L$ remain positively correlated.

In essence, extraction forces that activate once Ag/BCR clusters reach a critical size implement a physical form of negative feedback that is triggered upon completion of the proofreading process (assembly of a critical cluster). Nonlinearity of the force onset results in a stronger force per bond as a cluster grows in size (Fig.~5A, blue curve and inset), suppressing the dose dependence of the extraction level. We can understand the sensitivity to $L_0$ -- or the lack thereof -- from the cluster trajectories (Figs.~S6-S7) and distinctive features of the $\alpha_L\approx 0$ regime (Fig.~S8).
When more ligands are available, an inert force lacking negative feedback reduces the mean force per bond during contact formation (Fig.~S8C), allowing the cluster to reach a larger size before extraction occurs (Fig.~S8D) and to engage in more frequent rebinding during decline (Figs.~S6-S7). This results in a higher extraction level and a greater sensitivity to $L_0$. However, with a nonlinear cluster-sensing/adaptive force, negative feedback inhibits cluster growth while maintaining a stronger force per bond during contact formation when more ligands are present. This can at least partly explain the vanishingly small $\alpha_L$. 

Lastly, we demonstrate how variations in ligand concentration affect selection fidelity $\xi$. To mimic spatial inhomogeneity or temporal fluctuation of antigen abundance, we draw ligand quantity from a Gaussian distribution $\mathcal{N}(L_0, \sigma_L)$. As shown in Fig.~5C, under $F(t)$, fidelity falls as $\sigma_L$ increases, whereas discrimination performance remains high under $F(m_\mathrm{max})$. That is, an adaptive force can effectively
%maintain consistent fidelity of affinity discrimination, capable of 
buffer the influence of varying antigen concentration.

Therefore, dynamic forces that adjust to the clustering state provide the negative feedback needed to minimize the dependence on ligand quantity during extraction, thus enabling absolute discrimination in part of the parameter space.

\begin{figure*}[t]
\begin{center}
\includegraphics[width=0.95\textwidth]{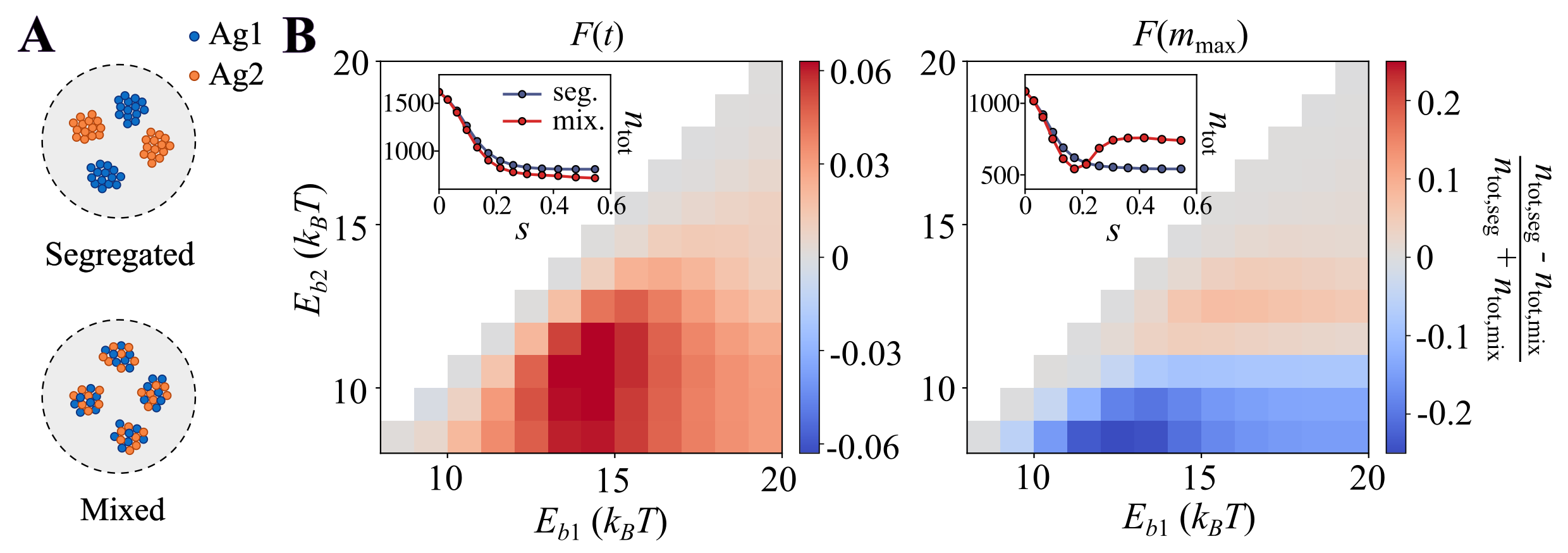}
\caption{\textbf{Preferred configuration of multi-type antigen presentation depends on B cell specificity.}
%{Different forcing schemes favor distinct spatial arrangement of multiple antigen types.}  
(A) We examine two extreme scenarios of multi-type antigen presentation: fully segregated (top) and well-mixed (bottom). (B) Heatmaps show the scaled difference in total antigen extraction per cell ($n_{\text{tot}} = n_{\text{Ag1}}+n_{\text{Ag2}}$) between segregated and mixed presentation scenarios. We assume every cell has 20 clusters with 100 antigens each.
Insets show the total extraction level in each scenario with respect to specificity $s = (E_{b1} - E_{b2}) / (E_{b1} + E_{b2})$. $s$ is varied by keeping $E_{b1}$ fixed at $17k_BT$ while varying $E_{b2}$ from $17k_BT$ to $5k_BT$. 
%The left panel is for a time-dependent force $F(t)$, and the right panel is for a cluster-sensing force $F(m_{\text{max}})$.
All results are averaged over 6000 independent runs.  
Under $F(t)$, segregation of different antigen types promotes extraction by specific and cross-reactive cells alike. Under $F(m_{\text{max}})$, mixing is preferred for specific cells, while segregation remains favorable for cross-reactive cells.  
Parameters: $k_{\text{on}} = 0.05\text{s}^{-1}$, $E_a = 12.6k_BT$, $F_0 = 600\text{pN}$, $t_c = 1\text{min}$, $m_c = 40$, $\beta = 5$.
}
\label{cross_reactive}
\end{center}
\end{figure*}

%\subsection{Antagonism effect due to coupling through force}
\subsection{Ligand antagonism due to load sharing}

Upon natural infection or vaccination, antigens are often presented in a mixture, comprising distinct mutants of the infecting virus or a cocktail of related antigenic variants. Researchers have demonstrated in experiments~\cite{schrader1975antagonism, kearney1976b, jameson1995t, ruppert1993effect} that both B and T cells exhibit ligand antagonism, the effect by which ligands below the activation threshold (antagonists) can alter the immune response to agonists. Antagonism can cause evasion of immune detection~\cite{klenerman1994cytotoxic, meier1995cytotoxic, schumacher2015neoantigens} and loss of vaccine efficacy~\cite{kent1997antagonism}. Conversely, understanding its mechanistic origin can help treat autoimmune diseases by antagonizing faulty cellular decisions~\cite{vanhove2017antagonist} and inform the design of multivalent vaccines.  

Although seemingly unrelated, absolute discrimination and ligand antagonism stem from a common root: the same negative feedback that flattens the dose-response curve also produces inhibitory coupling among receptors. In the context of antigen extraction, a shared load that couples the receptors can cause reduced agonist extraction, because antagonists with a slightly weaker affinity would trigger an additional load during cluster growth but cannot share the load in the dissociation stage.   

To detect antagonism of a physical origin, we examine agonist extraction in the presence of antagonists of varying affinities, contrasting $F(t)$ with $F(m_\mathrm{max})$. We quantify antagonism by the relative reduction in agonist extraction when mixed with antagonists (well-mixed in equal quantities):
%To investigate the interference between different antigen types, we consider a minimal model where two types of antigens (Ag1 and Ag2) with distinct off-rates are presented on APC in a well mixed form (see Sec.5 of SI for simulation details). B cells blindly bind to and extract two types of antigens simultaneously, while in the T cell selection step, only the foreign antigen (Ag1) is effective in terms of competing for T cell help. 
%Since different antigens within a cluster trigger and share the same loading force, acquisition of two types of antigens are coupled through force application. 
%To quantify the interference between different types of antigens, we define antagonism effect by the relative reduction in antigen extraction in the presence of a secondary antigen:
\[
&A_n \equiv \frac{\mu_{n}(E_{b1}) - \mu_{n}(E_{b1}, E_{b2})}{\mu_{n}(E_{b1})}, 
%\\ &A_{\tau} \equiv \frac{ \tau(k_{b1}) - \tau(k_{b1}, k_{b2}) }{\tau(k_{b1})},
\label{An}
\]
where $\mu_{n}(E_{b1})$ and $\mu_{n}(E_{b1}, E_{b2})$ denote respectively the mean extraction level of the agonists (of affinity $E_{b1}$) in the absence and presence of antagonists (of affinity $E_{b2}$). Without antagonism, $A_n=0$.
Collecting statistics from simulated extraction trajectories, we find that two force schemes yield opposing behaviors (Fig.~6C): $F(t)$ leads to largely cooperative interactions (negative antagonism) between ligands. By contrast, extraction under $F(m_\mathrm{max})$ exhibits significant antagonism; with a nonlinear force onset (colored curves), maximum antagonism occurs at intermediate antagonist affinities, whereas under a linear force (gray line), weak antagonists most severely inhibit agonist extraction. This behavior is analogous to that of artificial neural networks performing ligand classification under adversarial perturbations~\cite{rademaker2019attack}.

The contrast between force schemes can be understood from how ligand mixing alters the maximum attainable agonist cluster size $m_{1,\mathrm{max}}$ (Fig.~6A) and influences the mean rupture force per agonist bond $f_{r1}$ (Fig.~6B). Under $F(t)$, agonists can reach a larger cluster size as antagonists share the load and support further cluster growth. However, under $F(m_\mathrm{max})$, antagonists trigger an increase in force magnitude, reducing the agonist cluster. In both cases, mean rupture force per agonist bond peaks when antagonists have subthreshold affinities ($E_{b2}\leq E_{b1}$). In this regime, both ligands contribute to cluster growth and trigger a strong total force, but toward dissociation, antagonist bonds break sooner, leaving agonists alone to bear a high rupture force per bond (see Fig.~S9 for cluster trajectories). Much lower or higher antagonist affinity (relative to agonists) leads to weaker antagonism. When $E_{b2}\ll E_{b1}$, antagonist bonds break so frequently that they hardly affect the rupture force on agonist bonds. When $E_{b2}\geq E_{b1}$, two ligands become indistinguishable and break concomitantly toward dissociation.   

The combined effects explain the extraction outcomes (Fig.~6C). A smaller agonist cluster combined with a stronger rupture force per bond would result in lower agonist extraction, manifested as antagonism ($A_n>0$) under an adaptive force. To the contrary, with an inert force, an enlarged agonist cluster outweighs an increased rupture force, leading to ligand cooperativity ($A_n<0$). Fig.~S10 shows the underlying statistics.

An intriguing implication of this result is that extracting antigens under a shared load provides a mechanism for selecting against self-reactive B cells. Subject to a nonlinear adaptive force, loss of self-reactivity due to BCR mutation leads to reduced antagonism, which results in increased extraction of foreign antigens and hence acquisition of stronger survival signals from T helper cells. This could explain clonal redemption, documented by experiment in both mice and humans~\cite{sabouri2014redemption, reed2016clonal}, which posits that dual self-/foreign-binding B cells can be rescued by mutation away from self-reactivity~\cite{burnett2019clonal}. Our finding suggests that negative selection of auto-reactive B cells might be accomplished through positive selection of B cells that fail to bind self-Ag, since mutating away from self-binding can enhance foreign-binding by reducing antagonism.   
% rapid selection for mutations that decrease affinity for the self-Ag is followed by slower improvement in affinity to its foreign counterpart
This in turn reinforces a picture of GC positive selection centered on T helper cells.

\subsection{Preferred configuration of multi-type antigen presentation}

We finally assess how spatial organization of multiple antigen types influences total extraction. This is needed for understanding immune responses to 
%multi-dose or cocktail 
multivalent vaccines, where all variant antigens are foreign. We anticipate that sorting different types into separate clusters may enhance or reduce overall extraction depending on forcing dynamics. 

Consider extreme scenarios of multi-type antigen presentation (Fig.~7A): completely segregated and perfectly mixed, for the same total dose. In Fig.~7B, we present the relative efficiency (scaled difference) in total extraction between two scenarios for B cells with varying antigen specificity; see Fig.~S11 for absolute levels of total extraction.
%[Red/Blue indicates that segregation/mixing is preferred.] 
Two force schemes show strongly differing outcomes. Under $F(t)$, whether a B cell is specific or cross-reactive, a well-mixed presentation reduces extraction. This follows from the fact that mixing antigens leads to a lower rupture force on weaker antigens and a much higher rupture force on stronger antigens, with the latter dominating due to nonlinear force effects. Under $F(m_\mathrm{max})$, however, the favorable spatial arrangement depends on specificity. If a B cell is cross-reactive ($E_{b1}\approx E_{b2}$), a segregated presentation is preferred, like under $F(t)$. For specific cells ($E_{b1}\gg E_{b2}$), however, mixing antigens results in greater total extraction. As shown in Fig.~S12, stronger antigens are visited more frequently (panel J) and experience a reduced rupture force (panel K) when mixed with weaker antigens, hence increased extraction (panel L). Therefore, although weaker antigens are barely extracted in either case, they aid in distributing stronger antigens into smaller clusters and regulating force dynamics to allow more efficient engagement with BCRs.  

Combined, these findings indicate that spatial arrangement of multiple antigen types may strongly affect the overall extraction efficiency for cells of varying specificity. Under adaptive forces, segregated presentation is favorable for cross-reactive cells, whereas mixing is preferred for specific cells. If force is inert instead,
%one can select against specific clones using segregated Ag presentation. Otherwise, 
demixing of different types would promote extraction by specific and cross-reactive cells alike.

\section{Discussion}

Through proofreading reactions, nonequilibrium drives allow a strong reduction of errors in biological recognition.
%by introducing delays.
Here we propose to generalize proofreading to cases where the drive and the effect occur on vastly different scales: force-generating assemblies inside the cell can control selective expansion of cell lineages over many generations. Like conventional KPR, MPR relies on kinetic complexity of reaction pathways to be effective. Unique to MPR, the ability of intracellular forces to pattern the cell-cell interface is exploited to 
propagate information across scales.
%adaptively modulate coupled kinetics of competing pathways leading to a collective readout.
%(responsible for contact formation and dissociation). 
%[In other words, rather than amplify free energy differences between individual substrates like in conventional KPR, adaptive MPR enhances distinguishability of nearby distributions in parameter space, thus ensuring reliable ranking based on noisy readouts.] 
Although we demonstrate MPR through clonal selection in immune response, given the ubiquity of contractile force ratchet and force-sensitive receptor-ligand species, we expect this mechanism to operate in broad contexts, where cells communicate through physical engagement.

We examine the capacity of MPR to enhance selection fidelity by considering antigen extraction under dynamic forces. We show that elucidating the form and effect of the mechanical feedback %between coupled kinetics of load-sharing bonds and force exertion according to the clustering state 
is crucial for understanding the basis of clonal selection. This is because the mapping from affinity excess of receptors to competitive advantage of clones is strongly shaped by force dynamics. Both an inert force that activates after a preset delay and an adaptive force that adjusts to the clustering state are biologically plausible and supported by experimental data (Fig.~1). 
%Moreover, both schemes recapitulate temporal data of antigen extraction as well as in vivo distributions of contact duration. 
Interestingly, our model identifies an unambiguous indicator of mechanical feedback: 
%,namely, the sign of the scaling between contact duration and antigen abundance; 
adaptive forces allow faster detection of larger antigen abundance, opposite to inert forces (Fig.~2). 
In reality, two schemes may well coexist, with one being dominant depending on conditions, making MPR a flexible mechanism. 

By quantifying the contribution from different stages of force-driven extraction, we identify affinity-dependent discriminatory regimes and associated optimal force schemes (Fig.~3). At low BCR affinities, a ramping force that guides cluster growth toward the bifurcation point is the key to optimizing discrimination. At high affinities, a step force that closes the affinity gap between BCR and tether bonds maximizes fidelity. In addition, an optimal balance of speed and accuracy is achieved with an intermediate delay in force onset -- a further delay cannot improve accuracy (Fig.~4). Note that faster extraction of antigen allows efficient accumulation of survival signals and, in turn, stronger fitness gain, which may further enhance the fidelity of clonal selection across generations. 

An intriguing feature of the MPR mechanism presented here is ligand antagonism (Fig.~6). Conceptually, the combination of a sequence of assembly reactions with a negative feedback (here, supplied by nonlinear adaptive forces) produces inhibitory coupling between agonists and non-agonists, in close analogy to the biochemical counterpart in early activation of T lymphocytes~\cite{Francois:2013}. It will be interesting to explore a unifying framework of biological recognition that generalizes the concept of antagonism in light of adversarial perturbations (structured variations in ligand background)~\cite{rademaker2019attack}. Furthermore, our results have practical implications for autoimmunity and multivalent vaccines. We find that sub-threshold non-agonists cause maximum antagonism. A key implication is that negative selection of auto-reactive B cells can be achieved by positive selection of cells that fail to bind self-antigens, since mutating away from self-binding can enhance foreign-binding by reducing antagonism. On the other hand, our model predicts that under adaptive forces, presenting distinct antigens in separate clusters may increase the overall extraction by cross-reactive cells but reduce extraction by specific cells (Fig.~7). Thus, one can manipulate response specificity by patterning vaccine antigens on the presenting surface (e.g. a nanoparticle). 

One might expect that the need to maintain clonal diversity would compromise selection fidelity, as it likely caps the expansion of high affinity clones to make room for low affinity ones. Our results suggest that MPR may offer a way around this potential tradeoff. As we showed earlier~\cite{jiang2023molecular}, force exertion via a tug-of-war configuration reveals additional internal degrees of freedom of the BCR-Ag-tether complex, allowing a wide variety of binding phenotypes to coexist. 
%This in turn maintains phenotypic diversity among coevolving B cell lineages. 
The analysis here further indicates that the gain in diversity via a physical route is not incompatible with faithful selection. Instead, exertion of pattern-sensing dynamic forces can promote robust amplification of potent clones (against the current challenge) while keeping the phenotype spectrum broad (for coverage of future threats). 

Furthermore, we argue that the adaptive immune system likely deploys complementary kinetic and mechanical proofreading mechanisms to secure information transfer at different developmental stages or organization levels of response. A recent theoretical work~\cite{moran2024nonequilibrium} proposed that, in the face of exponentially replicating antigens, KPR is necessary for selective expansion of high-affinity naive B cells as input for affinity maturation. Here, we show that MPR may serve as the dominant mode during affinity maturation to sustain the selection pressure as affinity improves. Indeed, dramatic changes in contact pattern and force usage from naive B cells to GC B cells hint at a mode switch~\cite{nowosad2016, kwak2018}. 
%We will substantiate this perspective of seeing adaptive immunity as an epigenetic information system in future work.  

Perhaps most remarkably, enhancing the fidelity of clone selection via mechanical feedback yields desirable functional properties that appear incompatible and are not directly selected for: Negative feedback under adaptive MPR buffers against variations in ligand concentration (Fig.~5); meanwhile, this feedback allows tunable interactions between distinct ligands through the choices of relative affinity (Fig.~6) or spatial arrangement (Fig.~7). This led us to postulate that MPR might have evolved through selection for clonal expansion: MPR enhances ranking fidelity and hence increases the collective reproductive fitness of selected clones. The resulting boost in population growth, in turn, promotes the expansion of clones that possess MPR. Once mechanical feedback emerges, the functionalities it carries can add further selection pressure to maintain that feedback. An interesting avenue for future work is to describe clonal dynamics driven by the emergence, expansion, and propagation of mechanical feedback, setting the stage for uncovering an evolutionary origin of proofreading across scales.

\section*{Acknowledgments}
We thank Hongda Jiang for his essential inputs that made this work possible.
We are grateful for funding support from the National Science Foundation (NSF) Grant MCB-2225947 and an NSF CAREER Award PHY-2146581.

\bibliographystyle{unsrt}

\newpage

\bibliography{main}

\begin{thebibliography}{10}

\bibitem{ninio1975kinetic}
Jacques Ninio.
\newblock Kinetic amplification of enzyme discrimination.
\newblock {\em Biochimie}, 57(5):587--595, 1975.

\bibitem{hopfield1974kinetic}
John~J Hopfield.
\newblock Kinetic proofreading: a new mechanism for reducing errors in biosynthetic processes requiring high specificity.
\newblock {\em Proceedings of the National Academy of Sciences}, 71(10):4135--4139, 1974.

\bibitem{ehrenberg1980thermodynamic}
M~Ehrenberg and C~Blomberg.
\newblock Thermodynamic constraints on kinetic proofreading in biosynthetic pathways.
\newblock {\em Biophysical journal}, 31(3):333--358, 1980.

\bibitem{murugan2014discriminatory}
Arvind Murugan, David~A Huse, and Stanislas Leibler.
\newblock Discriminatory proofreading regimes in nonequilibrium systems.
\newblock {\em Physical Review X}, 4(2):021016, 2014.

\bibitem{mckeithan1995kinetic}
Timothy~W McKeithan.
\newblock Kinetic proofreading in t-cell receptor signal transduction.
\newblock {\em Proceedings of the national academy of sciences}, 92(11):5042--5046, 1995.

\bibitem{phillips2020molecular}
Rob Phillips.
\newblock {\em The molecular switch: Signaling and Allostery}.
\newblock Princeton University Press, 2020.

\bibitem{natkanski2013}
Elizabeth Natkanski, Wing-Yiu Lee, Bhakti Mistry, Antonio Casal, Justin~E. Molloy, and Pavel Tolar.
\newblock B cells use mechanical energy to discriminate antigen affinities.
\newblock {\em Science}, 340(6140):1587--1590, 2013.

\bibitem{suzuki2009}
Kazuhiro Suzuki, Irina Grigorova, Tri~Giang Phan, Lisa~M. Kelly, and Jason~G. Cyster.
\newblock Visualizing b cell capture of cognate antigen from follicular dendritic cells.
\newblock {\em Journal of Experimental Medicine}, 206(7):1485--1493, 06 2009.

\bibitem{eisen1964variations}
Herman~N Eisen and Gregory~W Siskind.
\newblock Variations in affinities of antibodies during the immune response.
\newblock {\em Biochemistry}, 3(7):996--1008, 1964.

\bibitem{victora2012}
Gabriel~D. Victora and Michel~C. Nussenzweig.
\newblock Germinal centers.
\newblock {\em Annual Review of Immunology}, 30:429--457, 2012.

\bibitem{victora2022germinal}
Gabriel~D Victora and Michel~C Nussenzweig.
\newblock Germinal centers.
\newblock {\em Annual review of immunology}, 40(1):413--442, 2022.

\bibitem{batista2001}
Facundo~D. Batista, Dagmar Iber, and Michael~S. Neuberger.
\newblock B cells acquire antigen from target cells after synapse formation.
\newblock {\em Nature}, 411(6836):489--494, 2001.

\bibitem{mcarthur2025antigen}
Hannah~CW McArthur, Anna~T Bajur, Maro Iliopoulou, and Katelyn~M Spillane.
\newblock Antigen mobility regulates the dynamics and precision of antigen capture in the b cell immune synapse.
\newblock {\em Proceedings of the National Academy of Sciences}, 122(20):e2422528122, 2025.

\bibitem{lanzavecchia1985antigen}
Antonio Lanzavecchia.
\newblock Antigen-specific interaction between t and b cells.
\newblock {\em Nature}, 314(6011):537--539, 1985.

\bibitem{victora2010germinal}
Gabriel~D Victora, Tanja~A Schwickert, David~R Fooksman, Alice~O Kamphorst, Michael Meyer-Hermann, Michael~L Dustin, and Michel~C Nussenzweig.
\newblock Germinal center dynamics revealed by multiphoton microscopy with a photoactivatable fluorescent reporter.
\newblock {\em Cell}, 143(4):592--605, 2010.

\bibitem{jiang2023immune}
Hongda Jiang and Shenshen Wang.
\newblock Immune cells use active tugging forces to distinguish affinity and accelerate evolution.
\newblock {\em Proceedings of the National Academy of Sciences}, 120(11):e2213067120, 2023.

\bibitem{jiang2023molecular}
Hongda Jiang and Shenshen Wang.
\newblock Molecular tug of war reveals adaptive potential of an immune cell repertoire.
\newblock {\em Physical Review X}, 13(2):021022, 2023.

\bibitem{jiang2024physical}
Hongda Jiang and Shenshen Wang.
\newblock Physical extraction of antigen and information.
\newblock {\em Proceedings of the National Academy of Sciences}, 121(39):e2320537121, 2024.

\bibitem{Nowosad:2016}
Carla~R Nowosad, Katelyn~M Spillane, and Pavel Tolar.
\newblock Germinal center b cells recognize antigen through a specialized immune synapse architecture.
\newblock {\em Nature Immunology}, 17:870 EP --, 05 2016.

\bibitem{Wang:2018}
Junyi Wang, Feng Lin, Zhengpeng Wan, Xiaolin Sun, Yun Lu, Jianyong Huang, Fei Wang, Yingyue Zeng, Ying-Hua Chen, Yan Shi, Wenjie Zheng, Zhanguo Li, Chunyang Xiong, and Wanli Liu.
\newblock Profiling the origin, dynamics, and function of traction force in b cell activation.
\newblock {\em Science Signaling}, 11(542):eaai9192, 08 2018.

\bibitem{knevzevic2018active}
Milo{\v{s}} Kne{\v{z}}evi{\'c}, Hongda Jiang, and Shenshen Wang.
\newblock Active tuning of synaptic patterns enhances immune discrimination.
\newblock {\em Physical review letters}, 121(23):238101, 2018.

\bibitem{bell1978}
George~I. Bell.
\newblock Models for the specific adhesion of cells to cells.
\newblock {\em Science}, 200(4342):618--627, 1978.

\bibitem{mason2011tuning}
Frank~M Mason and Adam~C Martin.
\newblock Tuning cell shape change with contractile ratchets.
\newblock {\em Current opinion in genetics \& development}, 21(5):671--679, 2011.

\bibitem{komianos2018stochastic}
James~E Komianos and Garegin~A Papoian.
\newblock Stochastic ratcheting on a funneled energy landscape is necessary for highly efficient contractility of actomyosin force dipoles.
\newblock {\em Physical Review X}, 8(2):021006, 2018.

\bibitem{batista2000b}
Facundo~D Batista and Michael~S Neuberger.
\newblock B cells extract and present immobilized antigen: implications for affinity discrimination.
\newblock {\em The EMBO journal}, 2000.

\bibitem{allen2007}
Christopher D.~C. Allen, Takaharu Okada, H.~Lucy Tang, and Jason~G. Cyster.
\newblock Imaging of germinal center selection events during affinity maturation.
\newblock {\em Science}, 315(5811):528--531, 2007.

\bibitem{nowosad2016}
Carla~R. Nowosad, Katelyn~M. Spillane, and Pavel Tolar.
\newblock Germinal center b cells recognize antigen through a specialized immune synapse architecture.
\newblock {\em Nature Immunology}, 17(7):870--877, 2016.

\bibitem{kwak2018}
Kihyuck Kwak, Nicolas Quizon, Haewon Sohn, Avva Saniee, Javier Manzella-Lapeira, Prasida Holla, Joseph Brzostowski, Jinghua Lu, HengYi Xie, Chenguang Xu, Katelyn~M. Spillane, Pavel Tolar, and Susan~K. Pierce.
\newblock Intrinsic properties of human germinal center b cells set antigen affinity thresholds.
\newblock {\em Science Immunology}, 3(29):eaau6598, 2018.

\bibitem{spillane2017b}
Katelyn~M Spillane and Pavel Tolar.
\newblock B cell antigen extraction is regulated by physical properties of antigen presenting cells.
\newblock {\em Biophysical Journal}, 112(3):126a, 2017.

\bibitem{gitlin2014clonal}
Alexander~D Gitlin, Ziv Shulman, and Michel~C Nussenzweig.
\newblock Clonal selection in the germinal centre by regulated proliferation and hypermutation.
\newblock {\em Nature}, 509(7502):637--640, 2014.

\bibitem{wang2021theory}
Bin Wang and Olga~K Dudko.
\newblock A theory of synaptic transmission.
\newblock {\em Elife}, 10:e73585, 2021.

\bibitem{cui2018identifying}
Wenping Cui and Pankaj Mehta.
\newblock Identifying feasible operating regimes for early t-cell recognition: The speed, energy, accuracy trade-off in kinetic proofreading and adaptive sorting.
\newblock {\em PloS one}, 13(8):e0202331, 2018.

\bibitem{shoval2012evolutionary}
Oren Shoval, Hila Sheftel, Guy Shinar, Yuval Hart, Omer Ramote, Avi Mayo, Erez Dekel, Kathryn Kavanagh, and Uri Alon.
\newblock Evolutionary trade-offs, pareto optimality, and the geometry of phenotype space.
\newblock {\em Science}, 336(6085):1157--1160, 2012.

\bibitem{altan2005modeling}
Gr{\'e}goire Altan-Bonnet and Ronald~N Germain.
\newblock Modeling t cell antigen discrimination based on feedback control of digital erk responses.
\newblock {\em PLoS biology}, 3(11):e356, 2005.

\bibitem{feinerman2008}
Ofer Feinerman, Ronald~N. Germain, and Grégoire Altan-Bonnet.
\newblock Quantitative challenges in understanding ligand discrimination by αβ t cells.
\newblock {\em Molecular Immunology}, 45(3):619--631, 2008.
\newblock Special section: Theories and Modelling of T Cell Behaviour.

\bibitem{Francois:2013}
Paul Fran{\c c}ois, Guillaume Voisinne, Eric~D. Siggia, Gr{\'e}goire Altan-Bonnet, and Massimo Vergassola.
\newblock Phenotypic model for early t-cell activation displaying sensitivity, specificity, and antagonism.
\newblock {\em Proceedings of the National Academy of Sciences}, 110(10):E888, 03 2013.

\bibitem{Francois:2016a}
Paul Fran{\c c}ois and Gr{\'e}goire Altan-Bonnet.
\newblock The case for absolute ligand discrimination: Modeling information processing and decision by immune t cells.
\newblock {\em Journal of Statistical Physics}, 162(5):1130--1152, 2016.

\bibitem{martinez2023long}
Ana Martinez-Riano, Shenshen Wang, Stefan Boeing, Sophie Minoughan, Antonio Casal, Katelyn~M Spillane, Burkhard Ludewig, and Pavel Tolar.
\newblock Long-term retention of antigens in germinal centers is controlled by the spatial organization of the follicular dendritic cell network.
\newblock {\em Nature immunology}, 24(8):1281--1294, 2023.

\bibitem{schrader1975antagonism}
John~W Schrader.
\newblock Antagonism of b lymphocyte mitogenesis by anti-immunoglobulin antibody.
\newblock {\em The Journal of Immunology}, 115(2):323--326, 1975.

\bibitem{kearney1976b}
John~F Kearney, Max~D Cooper, and Alexander~R Lawton.
\newblock B lymphocyte differentiation induced by lipopolysaccharide: Iii. suppression of b cell maturation by anti-mouse immunoglobulin antibodies.
\newblock {\em The Journal of Immunology}, 116(6):1664--1668, 1976.

\bibitem{jameson1995t}
Stephen~C Jameson and Michael~J Bevan.
\newblock T cell receptor antagonists and partial agonists.
\newblock {\em Immunity}, 2(1):1--11, 1995.

\bibitem{ruppert1993effect}
Jorg Ruppert, Jeff Alexander, Ken Snoke, Mark Coggeshall, Elizabeth Herbert, Douglas McKenzie, Howard~M Grey, and Alessandro Sette.
\newblock Effect of t-cell receptor antagonism on interaction between t cells and antigen-presenting cells and on t-cell signaling events.
\newblock {\em Proceedings of the National Academy of Sciences}, 90(7):2671--2675, 1993.

\bibitem{klenerman1994cytotoxic}
Paul Klenerman, Sarah Rowland-Jones, Steve McAdam, Jon Edwards, Susan Daenke, David Lalloo, Britta K{\"o}ppe, William Rosenberg, Diana Boyd, Anne Edwards, et~al.
\newblock Cytotoxic t-cell activity antagonized by naturally occurring hiv-1 gag variants.
\newblock {\em Nature}, 369(6479):403--407, 1994.

\bibitem{meier1995cytotoxic}
Ute-Christiane Meier, Paul Klenerman, Philip Griffin, William James, Britta K{\"o}ppe, Brendan Larder, Andrew McMichael, and Rodney Phillips.
\newblock Cytotoxic t lymphocyte lysis inhibited by viable hiv mutants.
\newblock {\em Science}, 270(5240):1360--1362, 1995.

\bibitem{schumacher2015neoantigens}
Ton~N Schumacher and Robert~D Schreiber.
\newblock Neoantigens in cancer immunotherapy.
\newblock {\em Science}, 348(6230):69--74, 2015.

\bibitem{kent1997antagonism}
Stephen~J Kent, Philip~D Greenberg, Mark~C Hoffman, Robert~E Akridge, and M~Juliana McElrath.
\newblock Antagonism of vaccine-induced hiv-1-specific cd4+ t cells by primary hiv-1 infection: potential mechanism of vaccine failure.
\newblock {\em Journal of immunology (Baltimore, Md.: 1950)}, 158(2):807--815, 1997.

\bibitem{vanhove2017antagonist}
Bernard Vanhove, Nicolas Poirier, Fadi Fakhouri, Laetitia Laurent, Bert ’t Hart, Pedro~H Papotto, Luiz~V Rizzo, Masaaki Zaitsu, Fadi Issa, Kathryn Wood, et~al.
\newblock Antagonist anti-cd28 therapeutics for the treatment of autoimmune disorders.
\newblock {\em Antibodies}, 6(4):19, 2017.

\bibitem{rademaker2019attack}
Thomas~J Rademaker, Emmanuel Bengio, and Paul Fran{\c{c}}ois.
\newblock Attack and defense in cellular decision-making: lessons from machine learning.
\newblock {\em Physical Review X}, 9(3):031012, 2019.

\bibitem{sabouri2014redemption}
Zahra Sabouri, Peter Schofield, Keisuke Horikawa, Emily Spierings, David Kipling, Katrina~L Randall, David Langley, Brendan Roome, Rodrigo Vazquez-Lombardi, Romain Rouet, et~al.
\newblock Redemption of autoantibodies on anergic b cells by variable-region glycosylation and mutation away from self-reactivity.
\newblock {\em Proceedings of the National Academy of Sciences}, 111(25):E2567--E2575, 2014.

\bibitem{reed2016clonal}
Joanne~H Reed, Jennifer Jackson, Daniel Christ, and Christopher~C Goodnow.
\newblock Clonal redemption of autoantibodies by somatic hypermutation away from self-reactivity during human immunization.
\newblock {\em Journal of Experimental Medicine}, 213(7):1255--1265, 2016.

\bibitem{burnett2019clonal}
Deborah~L Burnett, Joanne~H Reed, Daniel Christ, and Christopher~C Goodnow.
\newblock Clonal redemption and clonal anergy as mechanisms to balance b cell tolerance and immunity.
\newblock {\em Immunological reviews}, 292(1):61--75, 2019.

\bibitem{moran2024nonequilibrium}
Roberto Mor{\'a}n-Tovar and Michael L{\"a}ssig.
\newblock Nonequilibrium antigen recognition during infections and vaccinations.
\newblock {\em Physical Review X}, 14(3):031026, 2024.

\end{thebibliography}

\end{document}